\documentclass[conference]{IEEEtran}
\usepackage{extarrows}
\usepackage{engord}
\usepackage{paralist}
\usepackage{subfig}
\usepackage{epstopdf}
\ifCLASSINFOpdf
 \usepackage{graphicx}
\else
\fi
\begin{document}
%
\title{Tamperproof IoT with Blockchain}


\author{\IEEEauthorblockN{Guangsheng Yu\IEEEauthorrefmark{1}, Ren Ping Liu\IEEEauthorrefmark{1}\IEEEmembership{Senior Member, IEEE}, J. Andrew Zhang\IEEEauthorrefmark{1}\IEEEmembership{Senior Member, IEEE}, Y. Jay Guo\IEEEauthorrefmark{1}\IEEEmembership{Fellow, IEEE}}
\IEEEauthorblockA{\IEEEauthorrefmark{1} Network Security Lab, Global Big Data Technologies Centre, University of Technology Sydney}
Email: \{guangsheng.yu, renping.liu, andrew.zhang, jay.guo\}@uts.edu.au
}
\maketitle

\begin{abstract}
We investigate the tamper-resistant property of Blockchain and its effectiveness for IoT systems. In particular, we implemented an IoT testbed, and built a Blockchain into the testbed. A number of tamper-resistance experiments were conducted and analyzed to corroborate the process of block validation in Blockchain. Our analysis and experimental results demonstrate the tamper-resistant capability of Blockchain in securing trust in IoT systems. The demonstration video is provided at [1].
\end{abstract}


%
\IEEEpeerreviewmaketitle

\section{Introduction}
The Internet of Things (IoT) is poised to transform our lives and unleash enormous economic benefit. However, inadequate data security and trust are seriously limiting its adoption. Blockchain, a distributed tamper-resistant database, has native resistance to data tamper by which  it  is  practically impossible to modify the data retroactively once a transaction has been recorded. We believe such tamper-resistance property can be of significant value in ensuring trust in IoT systems.

\; We aim at demonstrating the tamper-resistant capability of Blockchain in securing trust in IoT systems. An IoT testbed was constructed in our lab, and Ethereum based Blockchain was built in the testbed. A number of tamper-resistance experiments have been carried out and analyzed to examine the process of block validation in Blockchain on both full node and light node. Our demonstrations reveal that Blockchain has a dedicated tamper-resistant capability, which can be applied to IoT to ensure trusted data collection and sharing.
With this features used within P2P network, precisely by design, a Blockchain-based database can therefore constitute a trust-free decentralized consensus system. Note that trust-free means conventional \engordnumber{3} party as an arbitral body is filled in by common cryptographical theorems. This promotes Blockchain to be a suitable role for data recording, storage and identity management, especially for those sensitive data [2]. 

The connection between Blockchain and IoT has no longer been futuristic [4-9]. With the research running, scientists have handed out Blockchain-based IoT.
There is no doubt that Blockchain and IoT are the two hot tags in the field of science and technology. IoT, including sensors, vehicles and other moving objects, basically contains any use of embedded electronic components with the outside world communication equipment, in particular, it makes use of the IP protocol,
as shown in Fig. 1.
\begin{figure}[htbp]
\centering
  \includegraphics[height=2.8cm]{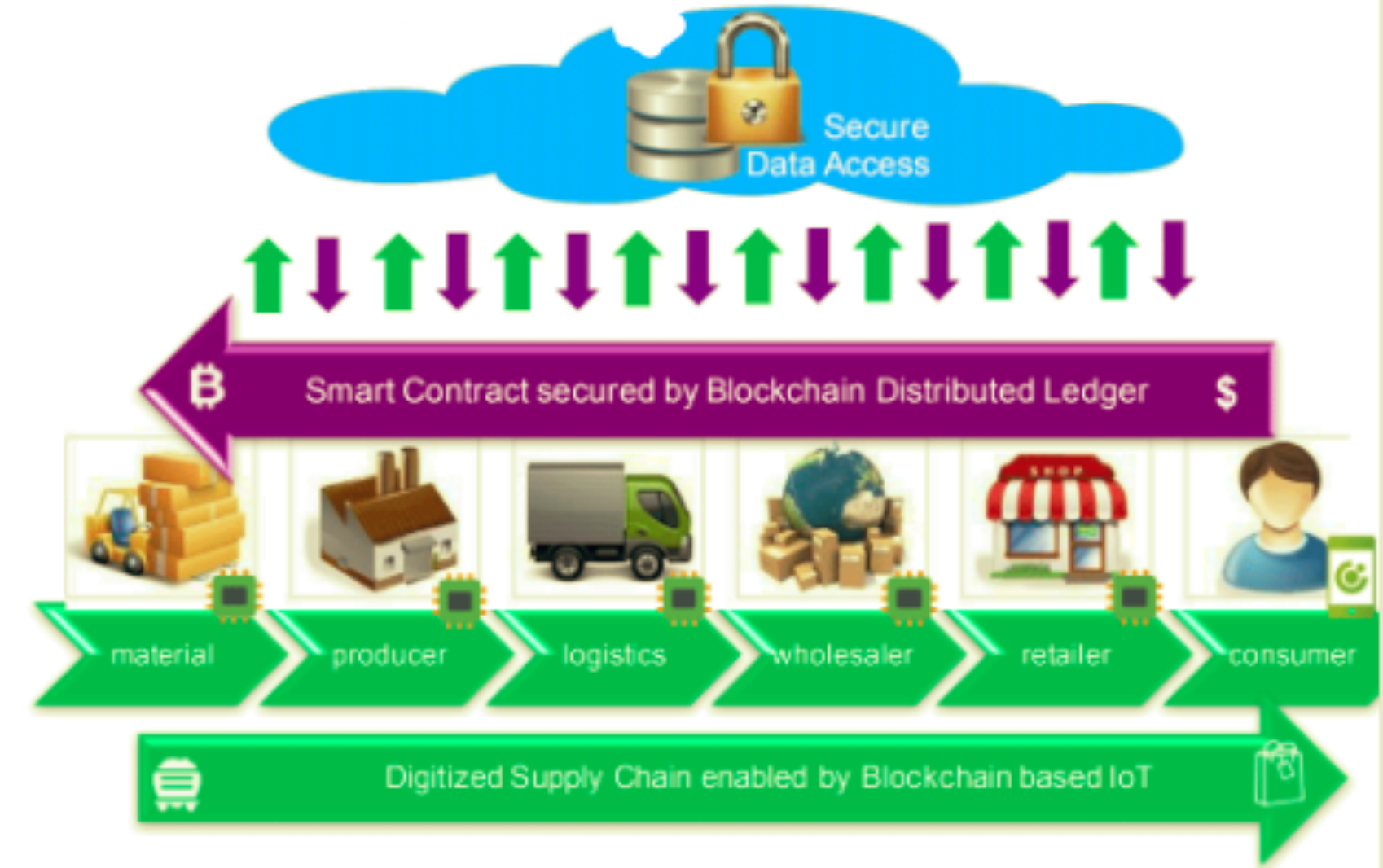}
  \caption{Trusted digital Supply Chain}
  \label{fgr:example}
\end{figure}

\section{Core Concept of Tamper-resistant}
\subsection{Key components}

We now explain the key components of tamper-resistant.

\; First of all, we point out the difference between the so-called 51\% attack and chain reorganization. Nodes will never import a block that fails validation. A 51\% attack consists of a 51\% of the miners forking off from a block in the past and creating a new chain that eventually beats the current canonical chain in total difficulty.

\; A reorganization is done only if a block being imported on a side fork  leads  to  a  higher  total  difficulty  on  that particular fork than the canonical fork. The blocks nonetheless still need to be valid. Note that triggering re-organization function is based on TotalDifficult in Proof-of-Work [2].

\; Synchronization happens all the time across each node starting from an identical genesis block in a chain. Each node runs a validation function to  validate each incoming block. No block will be accepted  unless passing the validation. 
\subsection{Validation procedures}
Throughout the validation procedures, the follow conditions are evaluated:
\begin{compactitem}
\item if \emph{StateRoot} $\in$ local levelDB, throw errors;
\item if \emph{ParentBlock} $\notin$ local levelDB, throw errors;
\item if \emph{StateRoot}$|$$_{\emph{ParentBlock}}$  $\notin$ local levelDB, throw errors;
\item if \emph{Validate(Header)} where \{\emph{nonce, difficulty, mixDigest\ldots}\} $\in$ \emph{Header} not passed, throw errors; 
\item if \emph{Validate(UncleHeader)} not passed, throw errors;
\item if \emph{Validate(GasUsed)} not passed $||$  \emph{Validate(bloom)} not passed, throw errors;
\item if \emph{TxHash} $\not=$ \emph{Hash(Txs)} $||$ \emph{ReceiptHash} $\not=$ \emph{Hash(Receipts)}, throw errors;
\item if \emph{StateRoot} $\not=$ 
\emph{StateRoot}$|$ $ \xLongleftarrow{Txs} \emph{CurrentStateRoot}$, throw errors. 
\end{compactitem}

  The last validation points out that since there is no transaction sent to those normal nodes, the state root after a state change will never equal to the state root of the header of the incoming block coming from an abnormal node.

A simple fraud of database does not account for any PoW computation. However, not only \emph{difficulty}, \emph{epochDataset} and \emph{mixhash}, but also the \emph{HeaderHash} is involved in calculating the targeting \emph{nonce}. It means arbitrary but sufficient amount of computation should be carried out on the tampered block. Even if sufficient amount of computation is satisfied, the record can be recovered by the canonical chain with the fastest speed effort, unless the hacker has control over 51\% computational power among the whole network.

\section{Tamper-resistance Demonstrations}
\subsection{System setup}
\subsubsection*{Hardware setup (shown in Fig.1)}
\begin{compactitem}
\item two workstations as mining nodes, shown in Fig.1(a);
\item three Raspberry Pi 3 B+, shown in Fig.1(b), attached with IoT sensors as end-point nodes, which are only allowed to look up and upload data without mining.
\begin{figure}[htb]
\centering
\subfloat[]{%
  \includegraphics[scale=0.17]{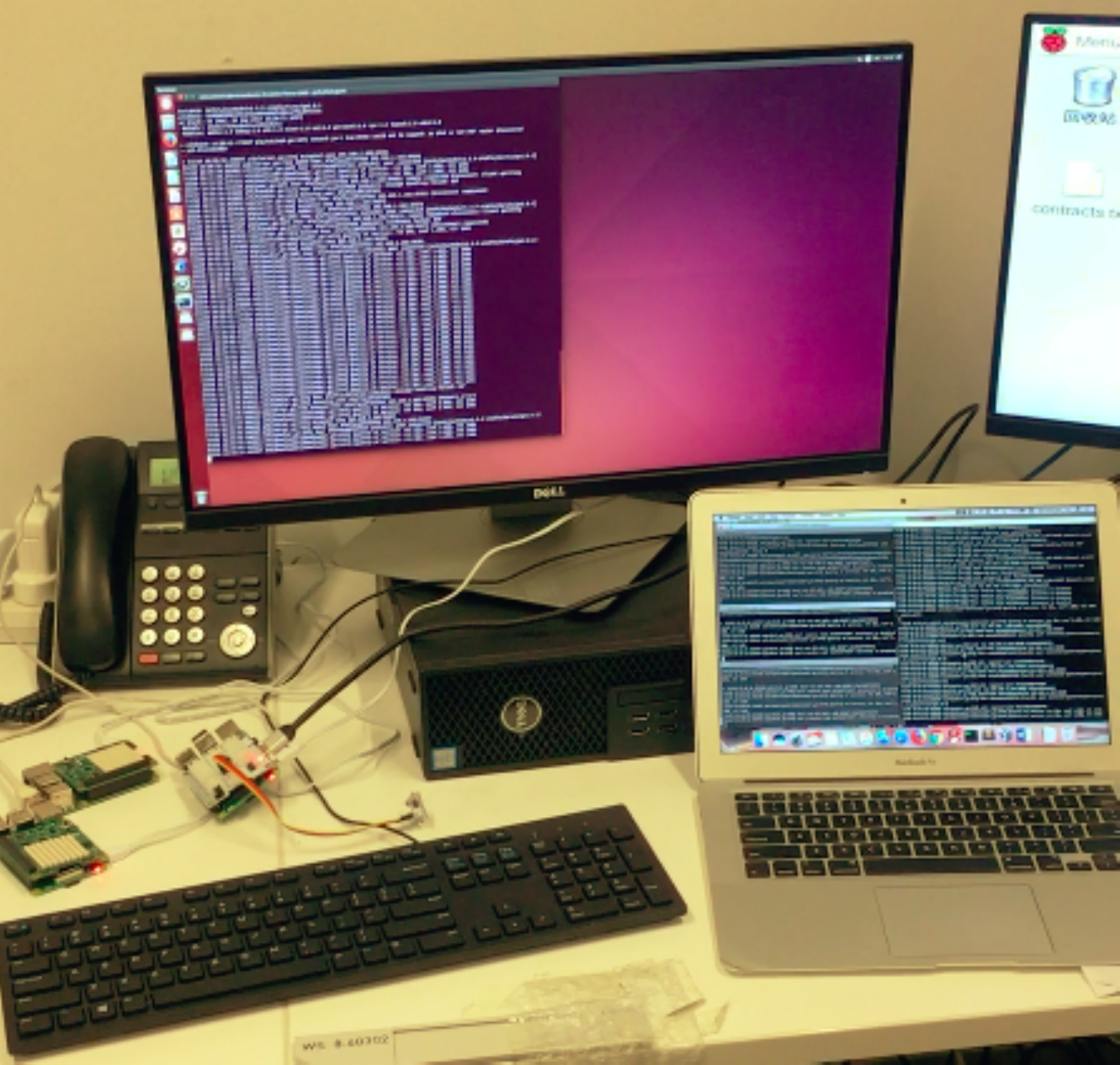}} \hfill
\subfloat[]{%
  \includegraphics[width=.24\textwidth]{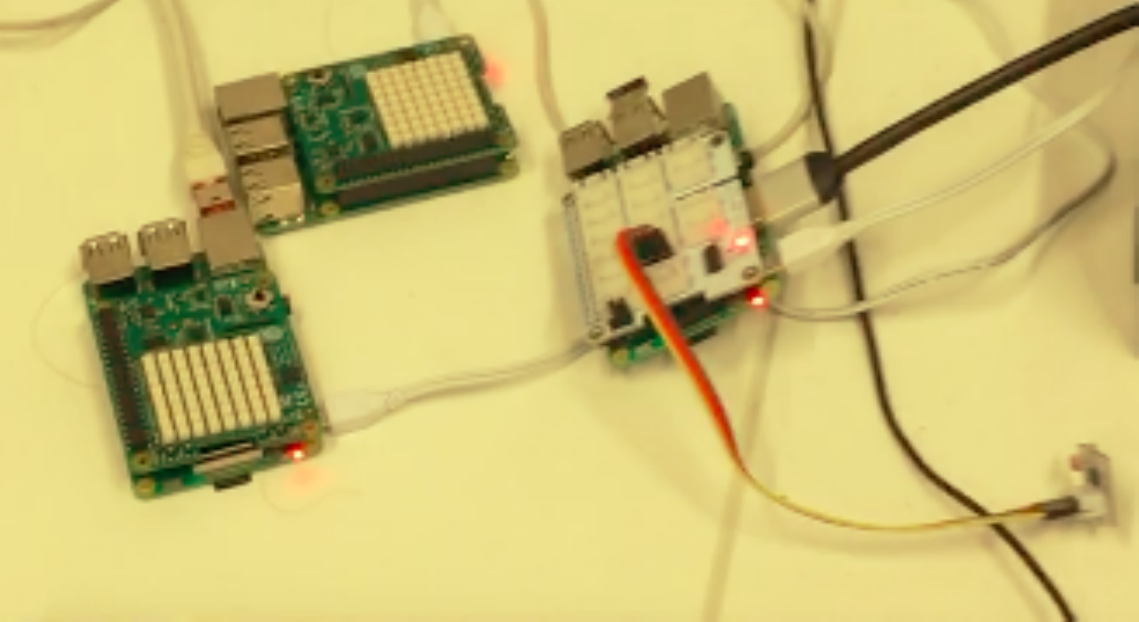}}
  \caption{Overview of testbed}
  \label{fgr:example}
\end{figure}
\end{compactitem}
\vspace{1mm}
\subsubsection*{Software setup}
\begin{compactitem}
\item Ubuntu 14.04 Trusty and Mac OS on mining nodes;
\item Raspbian on Raspberry Pi;
\item Golang Ethereum 1.5.9 for Blockchain [3];
\item Golang on hacking, tampering and logging;
\item Python on data receiving and encapsulation;
\item Javascript on Blockchain processing via web3 API [3].
\end{compactitem}
\vspace{2mm}
\subsubsection*{System Implementation}

\

The Raspberry Pi IoT devices equipped with temperature sensors measure room temperatures in the lab every 30 minutes. An Ethereum based Blockchain is built in the testbed. The IoT measurement data are encapsulated and uploaded to a pre-built contract in the Blockchain.
\subsection{Hacking scenarios and analysis}
We demonstrate several distinct scenarios where blocks are tampered with fake solution to PoW. Scenario analysis and experimental results are presented.
\vspace{1mm}
\noindent\emph{1) \ Non-mining node hacked:}
\

When a non-mining node is hacked, and the total difficulty is smaller than that of a normal block, the canonical chain always chooses the block with larger total difficulty in the context of a valid block. Thus this tampered block will be seen as an uncle block, the canonical chain will be recovered back by those normal blocks from other normal nodes.

\
When the total difficulty is greater than that of a normal block, other normal nodes will not accept the incoming blocks since the tampered block fails to pass the PoW validation during the synchronization. Once a future block whose total difficulty transcends that of the tampered block is generated by one   of the normal mining nodes, this tampered block will be seen as an uncle block, the canonical chain will be recovered back by those normal blocks from other normal nodes.

\vspace{1mm}
\noindent\emph{2) \ Mining node hacked:}
\

When a mining node hacked, and the total difficulty is smaller than that of a normal block, the results are the same as that of the non-mining node hacked case. Once this node starts mining, the uncle block will be broadcast at the same time for validation, contributing to errors throwing on other normal nodes shown as in Fig.2.
\begin{figure}[htbp]
\centering
  \includegraphics[scale=0.28]{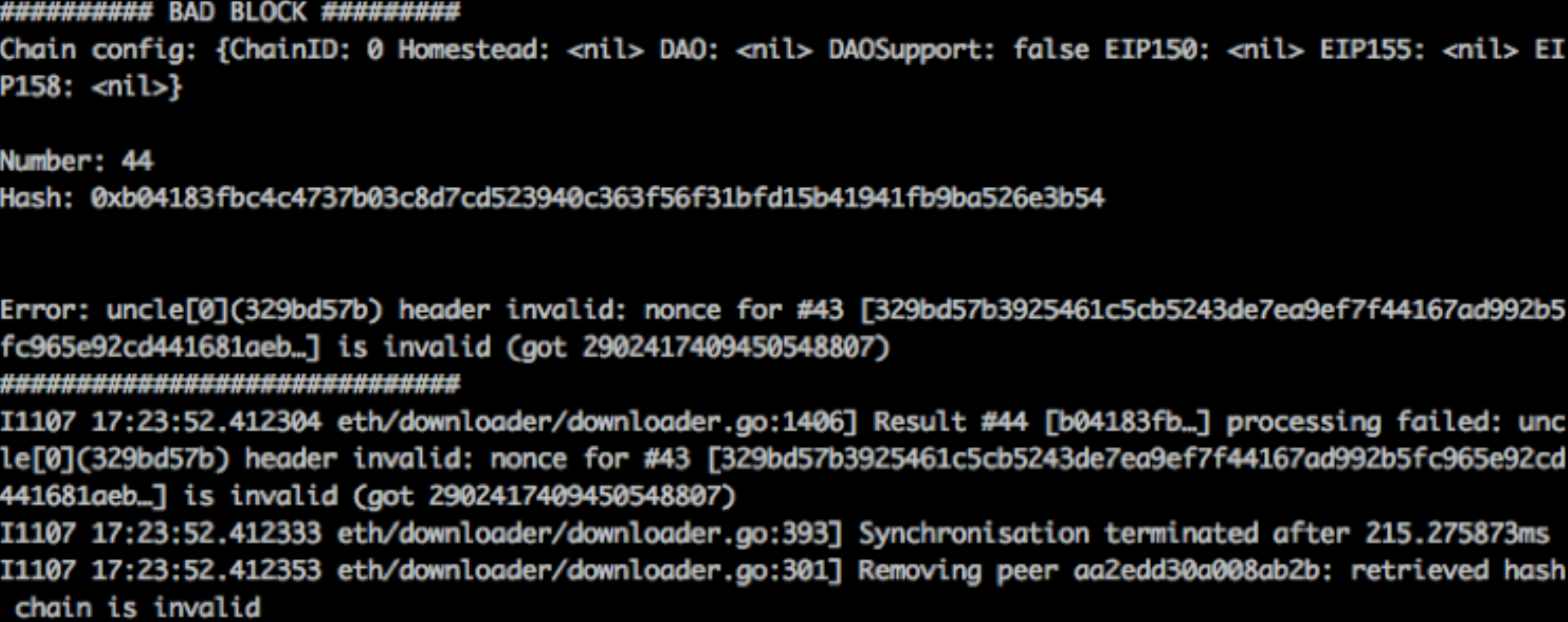}
  \caption{Bad Block with tampered uncle blocks}
  \label{fgr:example}
\end{figure}

\
When the total difficulty is greater than that of a normal block, it ends up being insufficient computational power. As a result, this tampered block will not be accepted by any other adversaries, and this mining node will be removed as a bad peer by other normal nodes, shown in Fig.3.
\begin{figure}[htbp]
\centering
  \includegraphics[height=1cm]{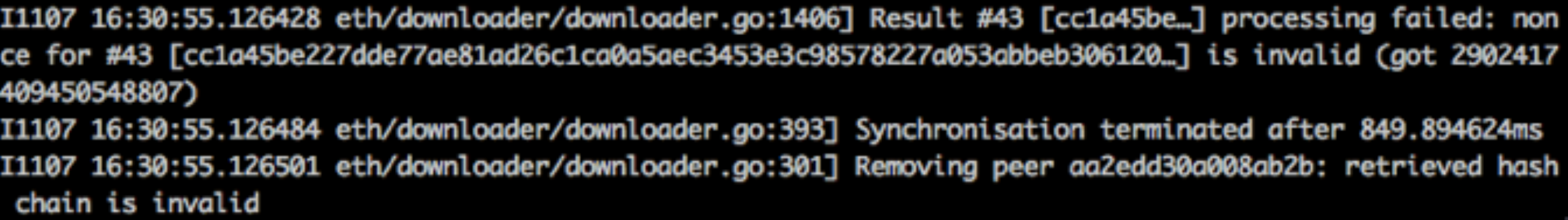}
  \caption{Insufficient computation leading to incorrect nonce}
  \label{fgr:example}
\end{figure}

\noindent\emph{3) \ Tampered Block on light node:}
\

We now investigate the scenario that a light node is hacked, and the PoW verification passes and the height of this light node is greater than that of those mining nodes. In this scenario, there will be no suitable peers available for this light node and any transactions  sent  from local will be broadcast to null, until the height is transcended by canonical chain, shown in Fig.4.

\begin{figure}[htb]
\centering
\subfloat[No suitable peer available when fetching data]{%
  \includegraphics[scale=0.29]{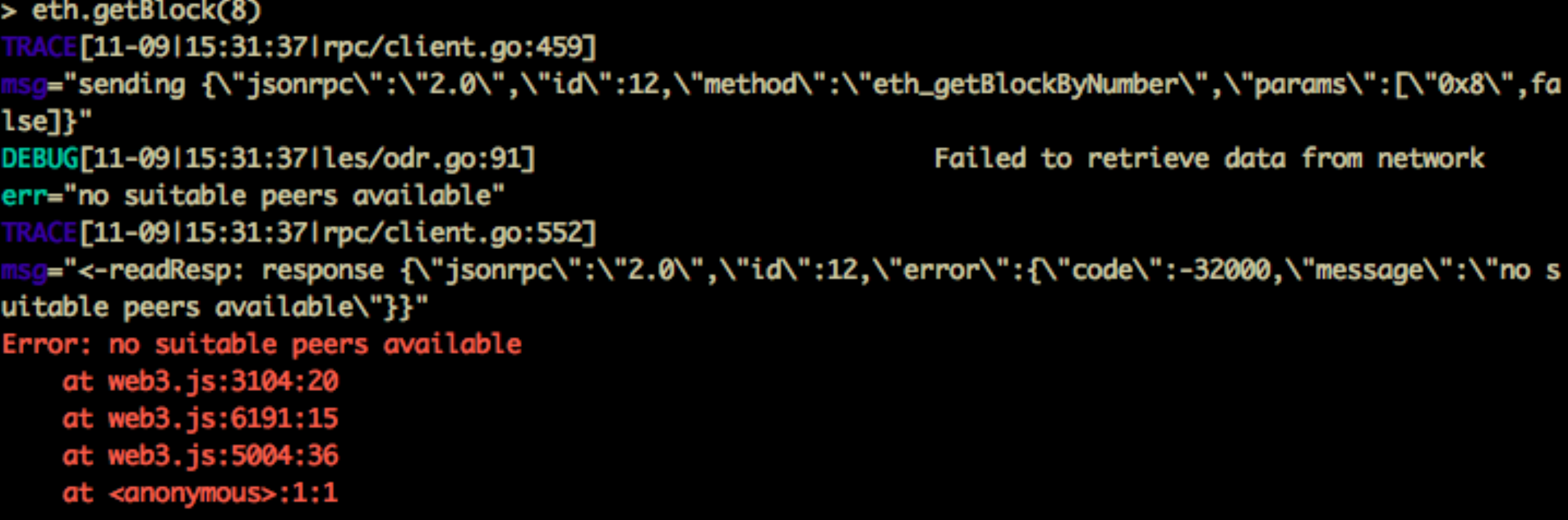}} 
  \\
\subfloat[No suitable peer available when sending transactions]{%
  \includegraphics[scale=0.3]{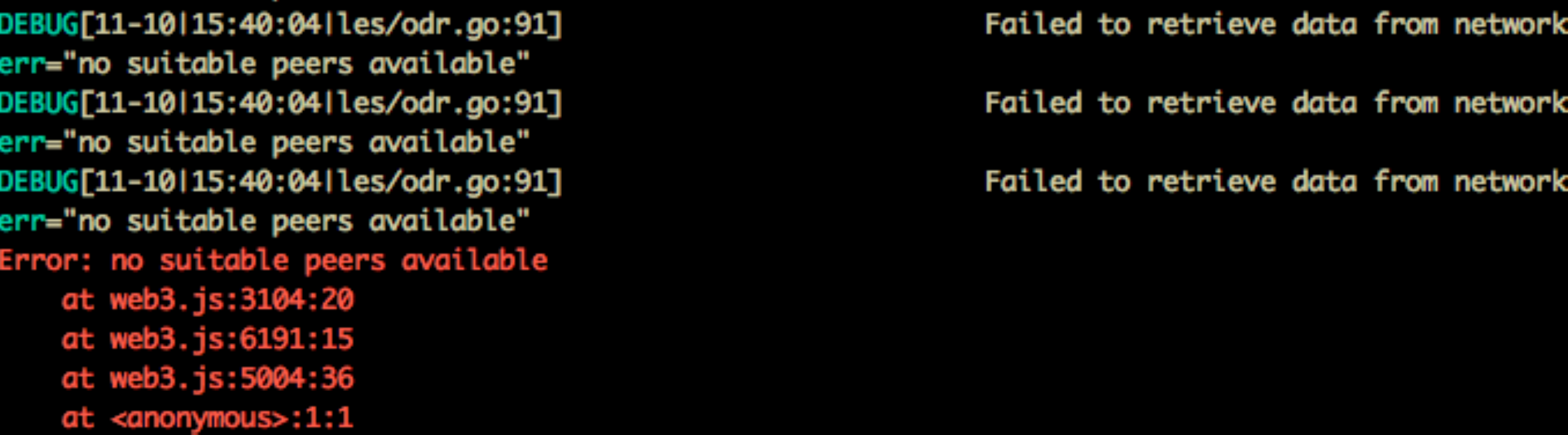}}
\end{figure}

\section{A Practical Demonstration}
\; We now demonstrate the effectiveness of the tamper-resistant property of Blockchain in protecting the IoT data records. In our Blockchain secured IoT testbed, one Raspberry Pi device was hacked, its temperature record was changed from the original measure of 34$^\circ$C to -4$^\circ$C. 
\par As soon as the tampering happened, the blockchain noticed this anomaly with a broken chain in the hacked node, which signals a tampering action.
\par Next, when the blockchain is synchronized, the abnormal block is  automatically  recovered  back  to the major one through the canonical chain. As a result the tampered record of -4$^\circ$C has been replaced by the original record of 34$^\circ$C. This is the chain reorganization process of the Blockchain. A log is generated to record which content had been changed unexpectedly, and this log is automatically uploaded onto Blockchain for future reference.
\par This demonstrated that the Blockchain can be applied to IoT to secure data records.






%

\end{document}